\documentclass[twocolumn]{article}
\usepackage[final]{graphics}
\usepackage{amsfonts}

\font\tenmsa=msam10
\font\sevenmsa=msam7
\font\fivemsa=msam5
\font\tenmsb=msbm10
\font\sevenmsb=msbm7
\font\fivemsb=msbm5
\newfam\msafam
\newfam\msbfam
\textfont\msafam=\tenmsa  \scriptfont\msafam=\sevenmsa
\scriptscriptfont\msafam=\fivemsa
\textfont\msbfam=\tenmsb  \scriptfont\msbfam=\sevenmsb
\scriptscriptfont\msbfam=\fivemsb

\global\mathchardef\lesssim "142E

\newcommand{\slL}{\raise.15ex\hbox{$/$}\kern-.53em\hbox{$L$}}
\newcommand{\slP}{\raise.15ex\hbox{$/$}\kern-.53em\hbox{$P$}}
\newcommand{\slR}{\raise.15ex\hbox{$/$}\kern-.53em\hbox{$R$}}
\newcommand{\slQ}{\raise.15ex\hbox{$/$}\kern-.53em\hbox{$Q$}}
\newcommand{\slK}{\raise.15ex\hbox{$/$}\kern-.53em\hbox{$K$}}
\newcommand{\slSigma}{\raise.15ex\hbox{$/$}\kern-.53em\hbox{$\Sigma$}}
\newcommand{\slcalP}{\raise.15ex\hbox{$/$}\kern-.63em\hbox{$\cal P$}}

\newcommand{\be}{\begin{equation}}
\newcommand{\ee}{\end{equation}}     
\newcommand{\bea}{\begin{eqnarray}}
\newcommand{\ena}{\end{eqnarray}}

\def\build#1\over#2{\mathrel{\mathop{\kern 0pt#1}\limits_{#2}}}

\font\tenimbf=cmmib10 at 10pt
\font\sevenimbf=cmmib10 at 7pt
\font\fiveimbf=cmmib10 at 5pt
\newfam\imbf
\textfont\imbf=\tenimbf
\scriptfont\imbf=\sevenimbf
\scriptscriptfont\imbf=\fiveimbf

\begin{document}
\date{April 2001}
\title{\bf{A simple out-of-equilibrium field theory formalism ?}}
\author{F.~Gelis$^{(1)}$, D.~Schiff$^{(2)}$, J.~Serreau$^{(2)}$}
\maketitle
\begin{center}
\begin{enumerate}
\item Brookhaven National Laboratory,\\
Physics Department, Nuclear Theory,\\
Upton, NY-11973, USA
\item Laboratoire de Physique Th\'eorique,\\
B\^atiment 210, Universit\'e Paris-Sud,\\
91405 Orsay, France
\end{enumerate}
\end{center}

\begin{abstract}
In this paper, we determine a condition of applicability of a very
simple formalism for the calculation of a fast process taking place in a
non-equilibrium medium, in which the particle distribution functions are
frozen in time.

Indeed, a well known obstacle for the use of such a naive formalism is
the appearance of the so-called ``pinch singularities''. We argue that
these potentially dangerous terms can be regularized, and that they
are negligible if the characteristic time-scale of the process under
study is small compared to the relaxation time of the system.
\end{abstract} 
\vskip 4mm 
\centerline{\hfill LPT-ORSAY 01/31, BNL-NT-01/7}

\section{Introduction}
The real-time formulation \cite{Schwi1,BakshM1,Keldy1,LandsW1} of
thermal field theories encodes the interactions with the heat-bath
through distribution functions that appear explicitly in the propagators
of the fields. In thermal equilibrium, those distributions depend only
on energy, and are the Bose-Einstein or Fermi-Dirac distributions. 

This formalism has been thought to be also suitable for the study of
processes taking place in a non-equilibrated medium
\cite{Keldy1,ChouSHY1}, provided one replaces the equilibrium
statistical weights by other functions of energy reflecting the new
particle distributions. These functions are allowed to have a space-time
dependence, but all the statistical weights of a diagram are taken at
the same point in space-time, so that this coordinate is just a passive
parameter in the calculation of a diagram. Therefore, the Feynman rules
of this formalism remain formally similar to those of the initial
real-time formalism. Of course, such a naive formalism is not expected
to be universally valid to calculate out-of-equilibrium processes, but
should be reasonably good for processes that are fast compared to the
relaxation time of the medium. In other words, the approximation
made by evaluating all the distribution functions at the same time in a
diagram should be satisfactory if the distributions indeed change very
little over the typical time it takes for this process to take place.

As a starting point of the discussion, let us come back to the work of
Altherr and Seibert [6] who realized that this naive formalism is
plagued by singular terms that seem to prevent its use for any
practical purpose. More precisely, these terms are ill-defined
products of propagators, with poles ``pinching'' the real energy axis,
also known as ``pinch singularities''. These pinch terms cancel
exactly in equilibrium thanks to the Kubo-Martin-Schwinger
\cite{Kubo1,MartiS1} (KMS) identities.  However, the KMS relations,
which are an expression of the detailed balance principle, are not
satisfied out-of-equilibrium, and this is responsible for the
non-cancellation of the pinch singularities.

Altherr \cite{Althe6} then proposed to regularize the pinch terms by
using effective propagators on which a width has been resummed.
Indeed, this width moves the poles of the propagator away from the
real axis, and the pinch terms become finite. For a width $\Gamma$,
they yield contributions proportional to $1/\Gamma$. However, the
order of magnitude of these pinch contributions has not been
investigated in a systematic way.

In fact, subsequent work by Bedaque \cite{Bedaq1} and Niegawa
\cite{Niega9,Niega8} shows that pinch terms are clearly related to
neglecting space-time inhomogeneities. In a more sophisticated
formalism based on Baym-Kadanoff equations, the would-be pinch terms
are compensated by terms containing gradients of the distribution
functions.


One could therefore assess the validity of the naive formalism by
estimating the order of magnitude of the pinch terms, compared to the
ordinary terms. Large pinch contributions would indicate that the
simplification that led to this formalism is not applicable, and that
we are in a situation where the relaxation of the medium has a non
trivial interplay with the process one is calculating. On the other
hand, negligible pinch terms are an indication that this naive
formalism can be used.

Our strategy in this paper is as follows. We start from this naive
formalism, and perform a consistency check by estimating the relative
order of magnitude of the pinch terms on a very simple example. From
there, we obtain a condition for these pinch terms to be negligible in
front of the regular ones. This condition defines the domain of
validity of the naive formalism, and coincides with intuition: pinch
terms are negligible if the typical time scale of the process under
study is much smaller than the relaxation time. In other words, if
this condition is satisfied, the calculation of a fast process at a
given time can be performed with out-of-equilibrium statistical
weights that are ``frozen'' at that time and one is allowed to neglect
pinch terms.

\section{Out-of-equilibrium retarded-advanced formalism}
\label{sec:RA}
We start by a generalization of the closed-time-path (CTP) formalism
\cite{Schwi1,BakshM1,LandsW1} in which one replaces the Bose-Einstein
and Fermi-Dirac distributions by arbitrary functions
\cite{Keldy1,ChouSHY1}. In addition to having an energy dependence
different from the equilibrium one, these distribution functions may
depend on a space-time coordinate.
However, an extra simplification of the formalism we are using here is
that all the statistical weights of a given diagram are evaluated at
the {\sl same point} in space-time. In other words, all gradients
are neglected. For bosons\footnote{We consider bosonic fields in this
  paper for the purpose of definiteness, but our arguments are valid
  for any type of field.}, the four components of the free matrix
propagator in this formalism are:
\begin{eqnarray}
&&\!\!\!\!\!\!G_0^{++}(P,X)=\Delta_F(P)+2\pi n(|p_0|,X)\delta(P^2-m^2)\;,
\nonumber\\
&&\!\!\!\!\!\!G_0^{--}(P,X)=\Delta_F^*(P)+2\pi n(|p_0|,X)\delta(P^2-m^2)\;,
\nonumber\\
&&\!\!\!\!\!\!G_0^{+-}(P,X)=2\pi(\theta(-p_0)+n(|p_0|,X))\delta(P^2-m^2)\; ,
\nonumber\\
&&\!\!\!\!\!\!G_0^{++} - G_0^{+-} - G_0^{-+} + G_0^{--} = 0 \; , 
\end{eqnarray}
with $\Delta_F(P)\equiv i/(P^2-m^2+i\epsilon)$ and where $X$ denotes
the point in space-time where the distribution functions are
evaluated. In order to keep the notations compact, we have implicitly
assumed that the distributions are isotropic in momentum space since
we have not included a $\mathbf p$ dependence. Since all the
distribution functions will be evaluated at the same space-time point
in the calculation of Feynman diagrams, we drop the variable $X$ in
the following.

This formalism contains only these propagators and the vertices
corresponding to the fundamental interactions. It does not contain any
higher-order correlator reflecting the initial statistical
distribution. In \cite{BellaM2}, it has been shown that this is
achieved if one drops all the non Gaussian correlations coming from
the initial density operator.

In this formalism, the vertices are the same as those of statistical
equilibrium, i.e. $\lambda^{---}=-\lambda^{+++}\not=0$ while all the
other components are vanishing. Pinch singularities
appear as products of $\delta(P^2-m^2)$, or as products of ${\cal
  P}/(P^2-m^2)$ where ${\cal P}$ denotes the principal part. It is
worth noting that this formalism requires only the distribution
function $n(p_0)$ for positive arguments; it is customary to extend
its definition to the complete real axis by requiring
$n(-p_0)=-1-n(p_0)$ for bosons.

It is convenient to switch to the retarded-advanced basis
\cite{AurenB1,EijckKW1} where two of the components of the free
propagator are vanishing, and where pinch singularities appear as
products of retarded and advanced propagators with the same momentum.
The transformation that leads to this formalism is a linear
transformation of the $2\times 2$ matrix made of the $G_0^{\pm\,
  \pm}$:
\begin{equation}
{G}_0^{XY}(P)\equiv \sum_{a,b=\pm}U^{Xa}(P) U^{Yb}(-P) G_0^{ab}(P)\; ,
\end{equation}
where the capital indices $X,Y$ take the values $R$ or $A$, and where
$U(P)$ is a $2\times 2$ matrix specifying the transformation. A choice
leading to the retarded-advanced formalism is
\begin{equation}
U(P)\equiv\pmatrix{
1 & -1 \cr
-n(-p_0) & -n(p_0) \cr
}\; ,
\end{equation}
where the first row is $R$, the second row is $A$, the first column is
$+$ and the second column is $-$. After this transformation, the
matrix propagator becomes:
\begin{equation}
{G}_0^{XY}(P)=\pmatrix{
0 & G_0^A(P)\cr
G_0^R(P) & 0\cr
}\; ,
\end{equation}
with
\begin{eqnarray}
&& G_0^A(P)\equiv G_0^{++}(P)-G_0^{-+}(P)\; ,\nonumber\\
&& G_0^R(P)\equiv G_0^{++}(P)-G_0^{+-}(P)\; .
\end{eqnarray}
In this formalism, the vertices $\lambda^{X ... YZ}$ can be expressed in terms
of $\lambda^{+\cdots+}$ and of the distribution functions \cite{AurenB1,EijckKW1}. In particular 
$\lambda^{A\cdots A}=0$. However, 
contrary to what happens in equilibrium, we have $\lambda^{R\cdots R}\not=0$.
This is a consequence of the fact that the KMS relations do not hold
out of equilibrium.

\section{Resummation of a self-energy}
\label{sec:width}
As it stands now, the out-of-equilibrium retarded-advanced formalism
is plagued by pinch singularities. These pinch terms can be made
finite, as noted in \cite{Althe6}, by resumming a width on the
propagators. One can see this width as a purely mathematical device
introduced to make the results finite, but it makes more sense to
identify it with the usual collisional width of particles in a plasma.
For instance, if the theory under consideration were QCD, this width
would be of order $g^2T\ln(1/g)$.

The self-energy that one has to resum in order to include the width
has a peculiarity out-of-equilibrium: the component $\Sigma^{RR}$ do not
vanish, contrary to what happens in equilibrium. Indeed, the matrix
corresponding to the self-energy in the RA formalism is:
\begin{equation}
\Sigma^{XY}(P)=\pmatrix{
\Sigma^{RR}(P) & \Sigma^R(P)\cr
\Sigma^A(P) & 0 \cr
}\; .
\end{equation}
This can be checked explicitly by using the previous Feynman rules for
the retarded-advanced formalism, or can be understood by relating this
component to the more familiar CTP formalism, which is done by means
of the following relation:
\begin{equation}
\Sigma^{RR}(P)=n(p_0)\Sigma^{+-}(P)-(1+n(p_0))\Sigma^{-+}(P)\; .
\end{equation}
The right hand side of the previous equation is the usual collision
term that appears in the Boltzmann equation, and it is known not to
vanish out-of-equilibrium. Taking into account this extra component in
the resummation, the resummed propagator is \cite{Althe6,CarriDT1}:
\begin{equation}
G^{XY}(P)=\pmatrix{
0 & G^A(P)\cr
G^R(P) & -i\Sigma^{RR}(P) G^A(P) G^R(P)\cr
}\; ,
\end{equation}
where $G^A$ and $G^R$ are the resummed retarded and advanced
propagators:
\begin{eqnarray}
&& G^R(P)\equiv{{G_0^R(P)}\over{1+iG_0^R(P)\Sigma^R(P)}}\; ,
\nonumber\\
&& G^A(P)\equiv{{G_0^A(P)}\over{1+iG_0^A(P)\Sigma^A(P)}}\; .
\end{eqnarray}
The important property of the resummed matrix propagator is that its
$G^{AA}$ component is not zero out of equilibrium. Moreover, this
component is coming from the pinch terms, regularized by the imaginary
part of the self-energy $\Sigma$.

In the following, when we talk about the ``contribution of pinch
terms'', we have in mind the contribution of the $G^{AA}$ component of the
full propagator (which does not exist in equilibrium, and
could\footnote{Note that a product like $G^A(P)G^R(P)$ does not
  necessarily become infinite when the width $\Gamma\to 0$. It becomes
  singular only if the momentum $P$ can reach the mass shell, i.e. if
  the corresponding diagram exhibits a mass-shell singularity already
  in equilibrium.} explode if the width is going to zero).

\section{Order of magnitude of pinch terms}
\label{sec:pinch}
It is now interesting to compare the relative order of magnitude of the
regular components $G^R$ or $G^A$ and of the component $G^{AA}$ which
contains the pinch terms. We have for instance:
\begin{equation}
{{G^{AA}(P)}\over{G^R(P)}}=-i\Sigma^{RR}(P)G^A(P)=
{{\Sigma^{RR}(P)}\over{P^2-M^2-2ip^0 \Gamma}}\; ,
\end{equation}
where $\Gamma$ is the width introduced on the propagator by the previous
resummation, and $M^2 = m^2 + {\rm Re} \Sigma_R$ is the resummed mass squared.

At this stage, it is very easy to express the previous ratio in terms of
the various length scales of the problem. First, the width $\Gamma$ is a
collision rate, and its inverse is the mean free path of the particle in
the medium:
\begin{equation}
\Gamma \sim \lambda_{\rm mean}^{-1}\; .
\end{equation}
The virtuality $P^2-M^2$ can be related by the uncertainty principle to
the typical lifetime of the off-shell state of momentum $P$. We can
write:
\begin{equation}
P^2-M^2\sim p^0 \lambda_{\rm coh}^{-1}\; .
\end{equation}
Physically, the space-time scale $\lambda_{\rm coh}$ (usually called
coherence length) one can define with the typical virtuality of the
propagators inside a diagram is a measure of the typical time it takes
for the process under study to take place\cite{AurenGZ2,Gelis11}. 

We need also an estimate for $\Sigma^{RR}(P)$. This can be obtained
if we recall that $\Sigma^{RR}(P)$ is the collision term of a Boltzmann
equation for particles of energy $p_0$~\cite{LeBellac}:
\begin{equation}
p_0 {{dn(p_0,t)}\over{dt}} = \Sigma^{RR}(P)\; ,
\end{equation}
 so that we can write:
\begin{equation}
\Sigma^{RR}(P)\sim p_0 \lambda^{-1}_{\rm non\ eq}\; ,
\end{equation}
where $\lambda_{\rm non\ eq}$ is the scale characterizing
the relaxation of the medium.  Collecting everything, we find the
relative order of magnitude of the pinch terms to be
\begin{equation}
{{{\rm pinch}}\over{{\rm reg}}}\sim
{{\lambda^{-1}_{\rm non\ eq}}\over{{\rm Max}\Big(\lambda^{-1}_{\rm mean},
\lambda^{-1}_{\rm coh}\Big)}}
\; .
\label{eq:ratio}
\end{equation}

We can now discuss different limits. If the microscopic scale
$\lambda_{\rm coh}$ is the smallest of the three scales:
$\lambda_{\rm coh}\ll \lambda_{\rm mean},\lambda_{\rm non\ eq}$, we
have:
\begin{equation}
\left.{{{\rm pinch}}\over{{\rm reg}}}\right|_{{\rm small\ }\lambda_{\rm coh}}\sim
{{\lambda_{\rm coh}}\over{\lambda_{\rm non\ eq}}}\ll 1\; .
\end{equation}
In other words, the contribution of pinch terms is always negligible
if the process under consideration is much faster than the relaxation
of the medium.  Practically, when this is the case, one can just
discard the pinch terms.

On the contrary, when the coherence length is very large, this ratio
is instead
\begin{equation}
\left.{{{\rm pinch}}\over{{\rm reg}}}\right|_{{\rm large\ }\lambda_{\rm coh}}\sim
{{\lambda_{\rm mean}}\over{\lambda_{\rm non\ eq}}}\; .
\label{eq:slow}
\end{equation}
Usually, the remaining two scales are typically of the same order of
magnitude since it takes a few times the mean free path to equilibrate
the system.

We can now make a connection with the approaches of \cite{Bedaq1} and
\cite{Niega9,Niega8}, who investigated the interplay between pinch
singularities and the relaxation towards equilibrium.  We see that the
ratio of Eq.~(\ref{eq:ratio}) becomes infinite in the limit where
$\lambda_{\rm non\ eq}$ becomes small, which is precisely the limit
where effects of the relaxation are very important. This is equivalent
to saying that neglecting the relaxation in the bare formalism of
section \ref{sec:RA} is responsible of pinch singularities.  However,
we add the following precision: the importance of pinch terms depend
on the typical time-scale associated with the process under
consideration; and they are negligible whenever the process is much
faster than the relaxation. In other words, the calculation of such a
fast process at a given time can be done with formulas similar to the
equilibrium ones, with ``frozen'' out-of-equilibrium distributions.
In order to specify these distributions, one needs to solve an
appropriate Boltzmann equation with given initial conditions.

In the case of a very slow process (Eq.~(\ref{eq:slow})), one cannot
disentangle the evolution of the medium from the process under study,
and it becomes necessary to keep track of the gradients. In addition,
it becomes very problematic to define quantities like production
rates, since the usual formulas of thermal field theory give local
rates.

\section{Conclusions}
\label{sec:conclusions}
In this paper, we have investigated the consistency of a naive
out-of-equilibrium field theory formalism in which rates are
calculated at a given time, neglecting all gradients. We have estimated
the relative order of the pinch terms and found that that they are
always negligible for processes that are characterized by a time-scale
very short in front of the relaxation scale.

In other words, the naive simplification which consists in dropping
all the gradients and in using distribution functions that are
``frozen'' in time leads to a consistent formalism provided the
statistical weights have very small variations over the typical time
it takes for the process under study to take place.

On the contrary, the situation is much more involved if this condition
is not satisfied, i.e. for a process whose typical time-scale is
comparable to the relaxation time itself. That this situation is
difficult to handle should not come as a surprise since in this
regime a local kinetic theory is not applicable: one has to go back to
first principles and solve Baym-Kadanoff-type equations.

\section*{Acknowledgements}
We would like to thank R. Baier for useful discussions.  The work of
F.G. is supported by DOE under grant DE-AC02-98CH10886.  F.G. would
also like to thank the European Center for Theoretical Studies in
Nuclear Physics and related areas (ECT*, Trento) where part of this
work has been performed, for its hospitality and support.

\bibliographystyle{unsrt}

\end{document}